\documentclass[aps,pra,twocolumn,floatfix,superscriptaddress,nofootinbib,showpacs,a4paper,balancelastpage,twoside]{revtex4-2}
\usepackage{graphicx, amsmath,amssymb,psfrag,empheq,hyperref,natbib}
\usepackage{dcolumn}
\usepackage{bm}
\usepackage{epsfig}
\usepackage{xcolor}
\usepackage{amsfonts}
\usepackage{tabularx}
\usepackage{graphicx} 
\graphicspath{{Bilder/}{../Bilder/}}
\usepackage{wasysym}
\usepackage{float}
\usepackage{siunitx}
\usepackage{hyperref}
\renewcommand{\vec}[1]{\bm{#1}} 
\usepackage[paperwidth=210mm,paperheight=297mm,centering,hmargin=1.8cm,vmargin=2.5cm]{geometry}

\newcommand{\qbounce}{{\it{q}}{\sc{Bounce}}}				

\begin{document}
\title{Search for dark energy with neutron interferometry} 
\author{Hauke Fischer}
\affiliation{Atominstitut, Technische Universität Wien, Stadionallee 2, A-1020 Vienna, Austria}
\author{Christian Käding}
\affiliation{Atominstitut, Technische Universität Wien, Stadionallee 2, A-1020 Vienna, Austria}
\author{Hartmut Lemmel}
\affiliation{Atominstitut, Technische Universität Wien, Stadionallee 2, A-1020 Vienna, Austria}
\affiliation{Institut Laue Langevin, 38000 Grenoble, France}
\author{Stephan Sponar}
\affiliation{Atominstitut, Technische Universität Wien, Stadionallee 2, A-1020 Vienna, Austria}
\author{Mario Pitschmann}
\email{mario.pitschmann@tuwien.ac.at}
\affiliation{Atominstitut, Technische Universität Wien, Stadionallee 2, A-1020 Vienna, Austria}

%
\begin{abstract}
We use previously obtained experimental results by neutron interferometry to effectively constrain the parameter space of several prominent dark energy models. This investigation encompasses the environment-dependent dilaton field, a compelling contender for dark energy that emerges naturally within the strong coupling limit of string theory, alongside symmetron and chameleon fields. Our study presents substantial improvements over previous constraints of the dilaton and symmetron fields, improving parameter constraints by several orders of magnitude. However, the analysis does not yield any new constraints on the chameleon field. Furthermore, we establish constraints for the projected neutron split interferometer, which has recently concluded a decisive proof-of-principle demonstration. Our symmetron simulations reveal that depending on the parameter values there are multiple static solutions with increasing number of nodes and increasing energy inside a cylindrical vacuum chamber. This agrees with results obtained earlier in the literature for infinitely parallel plates. Interestingly, while these multiple solutions can correspond to domain walls forming inside the vacuum chamber, we also find solutions that do not reach their vacuum expectation value inside the vacuum chamber, but display multiple nodes nonetheless. 
\nopagebreak
\end{abstract}
%
%
\maketitle
%
\section{Introduction}

The origin of dark energy is one of the greatest unsolved problems in modern physics. Observations on the cosmic scale unveil an ongoing acceleration in the expansion of our Universe~\cite{SupernovaCosmologyProject:1997zqe, SupernovaSearchTeam:1998fmf, SupernovaSearchTeam:1998bnz}. An unknown substance, called dark energy, that fills the Universe is the most prominent explanation for this acceleration. While introducing a cosmological constant into the framework of general relativity can in principle explain an accelerated expansion, such an approach grapples with intricate issues of fine-tuning~\cite{Sola:2013gha}. For this reason, new hypothetical scalar fields have been postulated, which couple to gravity and can account for dark energy~\cite{Joyce:2014kja}. These scalar fields  typically induce so-called fifth forces. To avoid conflict with ongoing fifth force searches, they must exhibit some screening mechanism in order to evade detection. Several screening mechanisms have been theorized, including the chameleon~\cite{Khoury:2003rn,Khoury:2013tda}, K-mouflage~\cite{Brax:2012jr,Brax:2014wla}, Vainshtein~\cite{Vainshtein:1972sx} and Damour-Polyakov mechanisms~\cite{Damour:1994zq}.

The analysis in this article focuses on the dilaton~\cite{Brax:2010gi,Sakstein:2014jrq}, symmetron~\cite{Cronenberg:2018qxf,Brax:2017hna,Pitschmann:2020ejb} and chameleon field~\cite{burrage2018tests}. 

The exponentially decreasing potential of the dilaton is expected from the strong coupling limit of string theory~\cite{gasperini2001quintessence,damour2002violations,damour2002runaway} and has been suggested as a dark energy candidate. First, experimental constraints from Lunar Laser Ranging~\cite{muller2019lunar} and gravity resonance spectroscopy~\cite{abele2010ramsey,jenke2011realization} as well as projective constraints from an experiment measuring the Casimir effect~\cite{Sedmik:2021iaw} have been derived in~\cite{Brax:2022uyh,Fischer:2023koa}. Furthermore, \cite{Kading:2023mdk} has shown that additional constraints can be obtained by studying the dilaton-induced open quantum dynamics based on the formalism developed in \cite{Burrage:2018pyg,Burrage:2019szw,Kading:2022jjl}.
 Symmetrons employ spontanous symmetry breaking similar to the Standard Model Higgs~\cite{hinterbichler2010screening,hinterbichler2011symmetron}. In low density regions symmetrons can obtain a non-vanishing vacuum expectation value (VEV) leading to the presence of a fifth force, while the symmetry is restored in high density regions. Recently, symmetron fifth forces have even been suggested as alternative explanations for the observed effects otherwise attributed to particle dark matter \cite{Burrage:2016yjm,OHare:2018ayv,Burrage:2018zuj,Kading:2023hdb}. The symmetron field has already been constrained by several table top experiments, such as atomic interferometry~\cite{burrage2015probing,burrage2016using,burrage2016constraining,brax2016atomic},  E\"otwash experiments~\cite{upadhye2013symmetron}, with gravity resonance spectroscopy~\cite{Brax:2017hna,cronenberg2018acoustic,Pitschmann:2020ejb,Jenke:2020obe} and by atomic precision measurements~\cite{brax2023screened}.
 Chameleons~\cite{khoury2004chameleon} are known to screen by increasing their masses in dense environments and have already been investigated and constrained comprehensively~\cite{burrage2018tests} (see e.g. also~\cite{Ivanov:2016rfs}). Furthermore, \cite{Hartley:2019wzu} proposed that tightened constraints on both symmetrons and chameleons could be obtained from Bose-Einstein condensate (BEC) interferometry in the future. Similarly, it was suggested that popular scalar field models with screening mechanisms, including those discussed in the present article, could be investigated in BEC-based analogue gravity simulations \cite{Hartley:2018lvm}.

A common feature among these models is that they are screened in dense environments and that forces on large objects are suppressed since the fields only couple to a thin-shell below the surface of an object~\cite{burrage2018tests}. As a result, forces on macroscopic objects as e.g. in the Solar System are typically very weak, while small objects such as neutrons in a vacuum chamber are ideal tools to probe these fields. Neutrons propagating through a vacuum chamber would experience a significant phase shift, and the absence thereof can be used to set stringent constraints on these models. In~\cite{lemmel2015neutron,li2016neutron} neutron interferometry has already been used to constrain the chameleon model for a few parameters (albeit these constraint have been superseded~\cite{burrage2018tests}). We extend the analysis to the dilaton and symmetron field and to a larger part of the chameleon parameter space and derive improvements that can be made through a split crystal interferometer~\cite{lemmel2022neutron}. It should be noted that ultra-cold neutrons provide versatile tools in the search for new physics in addition to scalar fields with screening mechanisms, see e.g.~\cite{Abele:2015uua, Jenke:2019qkw, Pitschmann:2019boa, Sedmik:2019twj,Sponar21, Ivanov:2021bvk, Suda:2021pit, Muto:2022eok,Kading:2023cvp}. 

In section~\ref{sec:II}, we provide the theoretical background for the investigated scalar fields as well as the expected induced phase shift on neutrons in a vacuum chamber. Details on the experimental setup are given in section~\ref{sec:III}. This is followed by sections~\ref{sec:IV},~\ref{sec:V} and~\ref{sec:VI} describing the dilaton, symmetron and chameleon constraints, respectively. We conclude our findings in~\ref{sec:VII}. Finally, in appendices~\ref{sec:AppI} and~\ref{sec:AppII}, we provide details on the numerical solutions of differential equations and derive the phase shift formula, respectively.

\section{Theoretical background}\label{sec:II}

The effective potential of the scalar fields under consideration is given by 
\begin{align}
    V_{\text{eff}}(\phi;\rho) = V(\phi) + \rho A(\phi)\>,
\end{align}
where $V(\phi)$ describes the self-interactions of the field and the Weyl factor $A(\phi)$ couples to the ambient matter density $\rho$.
The dilaton (D), symmetron (S) and chameleon (C) models are defined by~\cite{burrage2018tests, Brax:2018iyo}
\begin{align}
    V_D(\phi) &= V_0\, e^{-\lambda_D \phi /m_{\text{Pl}}}\>,\\
    V_S(\phi) &= - \frac{\mu^2}{2}\,\phi^2 + \frac{\lambda_S}{4}\,\phi^4\>,\\
    V_C(\phi) &= \frac{\Lambda^{n+4}}{\phi^n}\>,
\end{align}
together with
\begin{align}
    A_D(\phi) &= 1 + \frac{A_2}{2} \frac{\phi^2}{m^2_{\text{Pl}}}\>,\\
    A_S(\phi) &= 1 + \frac{\phi^2}{2M^2}\>, \\
    A_C(\phi) &= e^{\phi / M_c} \simeq 1 + \frac{\phi}{M_c}\>.
\end{align}
The parameters of the dilaton are the energy scale $V_0$, related to the dark energy of the Universe, the dimensionless parameter $\lambda_D$ as well as the dimensionless coupling parameter $A_2$, while $m_{\text{Pl}}$ denotes the reduced Planck mass. The symmetron parameters are given by the tachyonic mass $\mu$, a dimensionless self-coupling constant $\lambda_S$ and $M$ as a coupling constant to matter with the dimension of mass. For chameleon models, $n \in \mathbb{Z}^+ \cup 2\mathbb{Z}^-\setminus$\{-2\} determines the power of the self-interaction potential, $\Lambda$ defines an energy scale that is sometimes related to the dark energy of the Universe and $M_c$ is a coupling constant with the dimension of mass.
To neglect possible couplings to matter of higher order, we restrict our analysis to
\begin{align}
    \frac{A_2}{2}\frac{\phi^2}{m^2_{\text{Pl}}}, \frac{\phi^2}{2 M^2}, \frac{\phi}{M_c} \ll 1\>, \label{cutoff}
\end{align}
which leads to sharp cut-offs in the displayed exclusion plots. For the chameleon field, this restriction is actually not necessary. However, regions where neutron interferometry is sensitive and this condition does not hold have already been constrained by other experiments.

The experimental setup is such that the neutron beam is split into two paths, in each the neutrons traverse a chamber containing some gas. For brevity, the chamber containing gas of significantly lower density will be denoted as "vacuum chamber", while the other chamber will be referred to as "filled with air". This leads to a relative phase difference as will be detailed in section~\ref{sec:III}. The presence of a scalar field inside a chamber induces an additional phase shift, which to leading order is given by
\begin{align}
    \delta\varphi_X = - \frac{m_n}{k_0} \int\limits_{\text{CFP}} U_X(\vec{x})\, ds\>,
\end{align}
with $X\in \{D, S, C\}$, the integration extending over the classical flight path (CFP) and 
\begin{align}
U_D(\vec{x}) &= \mathfrak{Q}_D\frac{A_2 m_n}{2 m^2_{\text{Pl}}}\left(\phi^2(\vec{x})-\phi^2_{\text{Air}}\right),\\
U_S(\vec{x}) &= \mathfrak{Q}_S\frac{m_n}{2 M^2}\left(\phi^2(\vec{x})-\phi^2_{\text{Air}}\right),\\
U_C(\vec{x}) &= \mathfrak{Q}_C\frac{m_n}{M_c}\left(\phi(\vec{x})-\phi_{\text{Air}}\right).
\end{align}
Here, $k_0$ is the wave number of the neutron, $m_n$ its mass, $\phi_{\text{Air}}$ is the field value that minimizes the potential in air and  $\mathfrak{Q}_X$ the screening charge of the neutron. 
We provide a derivation of these formulas and its applicability conditions in Appendix~\ref{sec:AppII}. Computing the field of the neutron and the chamber is in general a two-body problem between the neutron and the vacuum or air chamber. We approximate this problem by treating the neutron as a classical massive sphere with radius 0.5 fm in agreement with QCD (this approximation is refered to as "fermi-screening"), which allows the screening of the neutron to be taken into account approximately. Thereby, the potential is multiplied with a screening charge $\mathfrak{Q}$ that takes values between 0 (for fully screened spheres) and 1 (for unscreened spheres).
The analytical expressions used can be found in~\cite{Brax:2022uyh} for the dilaton and in~\cite{Brax:2017hna,Pitschmann:2020ejb} for the symmetron model. For the chameleon field, we derived a screening charge by following analogous reasoning as for the dilaton screening charge. We recall its derivation in Appendix~\ref{sec:AppIII}. In~\cite{Fischer:2023koa} a second approximation ("micron-screening") was considered, by extracting a radius of the neutron from the vertical extent of the wave function, which amounts to 5.9 $\mu$m in that experiment. The wave function in the neutron interferometer has a very particular shape, being extended over mm in one direction and nm in the others, cf. section~\ref{sec:III}. Hence we refrain from trying to extract a radius from it. However, it was verified that assuming a neutron radius larger than 0.5 fm, as could be extracted from any extent of the wave function, would lead to constraints containing the fermi-screening ones. Hence, fermi-screening constraints are more conservative.
\begin{figure*}[t]
\centering
\includegraphics[width=\textwidth]
{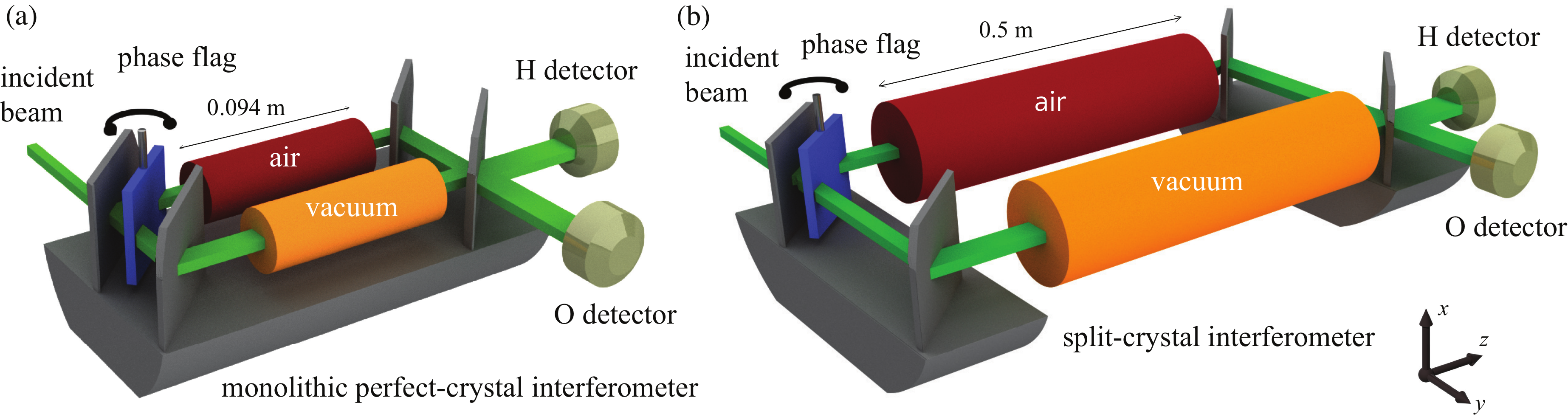}
\caption{Schematic illustration of neutron interferometric setups (not to scale) to test various prominent dark energy models using vacuum and air chambers in cylinder design. (a) Monolithic interferometer as used in ~\cite{lemmel2015neutron} and (b) proposed split-crystal interferometer setup with longer interaction region and larger beam separation, allowing for larger cylinder diameters. \label{fig:setup}}
\end{figure*}

For each field model it is necessary to compute $\phi$ inside the vacuum and  air chamber, which amounts to solving the non-linear differential equation
\begin{align}
    \Delta \phi (\vec{x}) = V_{\text{eff}, \phi} (\phi(\vec{x}); \rho(\vec{x}))\>. \label{eqn}
\end{align}
As boundary condition we demand that $\phi$ reaches its potential minimum $\phi_M$ inside the shell of the vacuum and air chamber, and restrict our analysis to parameters where this assumption is justified. Due to the absence of analytical solutions, the differential equations were solved numerically. Details about the algorithm employed are provided in Appendix~\ref{sec:AppI}.

Since every parameter combination requires a separate numerical solution, the computational cost of constraining a 3D parameter space in this way is unfeasible. 
Therefore, the rectangular vacuum and air chamber have been approximated by cylinders, which amounts to solving
\begin{align}
    &\left[\frac{1}{r}\frac{\partial}{\partial r}\left(r\frac{\partial}{\partial r}\right)  + \frac{\partial^2}{\partial z^2}\right]\phi(r,z)= V_{\text{eff}, \phi} (\phi; \rho)\>.\label{numeric}
\end{align}
These cylinders have the same length as the real chambers (which was 0.094 m in the experiment, 0.5 m was assumed for the split interferometer) and either just fit into the real chamber ($r =2$ cm in both chambers) or are just large enough to encompass the real chamber ($r = \sqrt{8}$ cm in the vacuum chamber and $r = \sqrt{3.25^2+2^2}$ cm in the air chamber). That the difference between both geometries is negligible has been verified and hence the final constraints have been computed assuming $r=2$ cm. 
For the future split interferometer a chamber with cylinder symmetry and radius $r = 4.75$ cm has been assumed. 

\subsection*{Constraint calculation profile mode}

In the experiment, two measurement modes were applied. First, in the profile mode, the pressure in the vacuum chamber was fixed to $10^{-4}$ mbar and the beam position varied. Initially, both beams passed through the center of either chamber. Next, the beam position was changed to a position closer to the chamber walls corresponding to a maximum displacement of $\sim$1.5 cm from the center. Since scalar fields are more suppressed close to the chamber walls than at the center of the chamber, the induced phase shift would be position dependent. The phase difference is measured for both positions, and the difference can be used to derive experimental constraints for scalar fields. The experiment constrains phase shifts exceeding 5.5 degrees at a confidence level of 95\%. The phase shift was computed using
\begin{align}
    \delta \varphi_{X,P} = - \frac{m_n}{k_0} \int_{-L/2}^{L/2}dz\,\big(U_X(0, z) -  U_X(0.015\, \text{m}, z)\big)\>,
\end{align}
for both the air and vacuum chamber. Parameter constraints are obtained by using
\begin{align}
    \big|\delta\varphi_{X,P}(\text{Air})-\delta\varphi_{X,P}(\text{Vacuum})\big| \geq 5.5^{\circ}\>.
\end{align}
Projective constraints for the split interferometer have been derived by the same equations and experimental parameters with the exception that the second beam position was assumed to be 4 cm displaced from the center.

\subsection*{Constraint calculation pressure mode}

In the pressure mode, both beams passed through the center of either chamber but the pressure inside the vacuum chamber was varied. The phase shift resulting from the highest pressure of $P_0=10^{-2}$ mbar was used as a reference and the pressure lowered to $P_1=2.4\times 10^{-4}$ mbar. This measurement mode  constrains phase shifts exceeding 3.5 degrees with a confidence level of 95\%. The vacuum chamber was simulated for both pressures and parameter constraints are obtained by using
\begin{align}
     \frac{m_n}{k_0}\bigg|\int_{-L/2}^{L/2}dz\,\big(U_{X,P_0}(0, z) - U_{X,P_1}(0, z)\big)\bigg| \geq 3.5^{\circ}\>.
\end{align}
The projective constraints for the split interferometer have been computed using the  same equations and experimental parameters.

\section{Experimental setup}\label{sec:III}

In a neutron interferometer~\cite{Rauch74, RauchBook} a beam of neutrons is split by amplitude division into two beam paths which are superposed coherently after passing through different regions of space (Mach-Zehnder geometry). One of the most straightforward experiments is the measurement of the neutron index of refraction of a sample material which allows to determine the coherent neutron scattering length of the nuclei contained in the sample. In this sense, the present experiment measures whether the index of refraction of vacuum indeed vanishes or not. 

The neutron beam has a typical diameter of a few mm to cm. The coherence length of the incoming neutrons is in the order of nm, so the beam consists of an incoherent ensemble of neutrons spread over the beam cross section. Each individual neutron, however, is coherently split into the two beam paths, separated by a few cm, and exhibits self-interference. Therefore, neutron interferometry has also been used to study the foundations of quantum mechanics, starting in the early 1970s with the
verification of the 4$\pi$ spinor symmetry~\cite{Rauch754Pi}, up to recent experiments on entanglement and weak values, reviewed e.g. in \cite{Sponar21}. 

Beam splitters and mirrors in neutron interferometry are based on Bragg diffraction on perfect silicon crystals in transmission (Laue) geometry. This implies that the neutron wave function is not only coherently split into two beam paths but also spread to a width of several mm within a single beam path \cite{RauchBook,rauch1996}. It is still subject of discussion whether the full extension of the wave function or the particle size of the neutron hast to be taken into account when calculating the scalar field around the neutron. As already pointed out, we have restricted our calculations to the latter since it gives the more conservative constraints.

The beam splitters of an interferometer have to be aligned to each other with nrad and sub-nm accuracy. Therefore, neutron interferometers are usually built monolithically out of a single crystal, cf. Fig.~\ref{fig:setup}~(a). This design allowed in our last experiment~\cite{lemmel2015neutron} for a beam separation of 50\,mm and a chamber length of 94\,mm. However, we would gain in sensitivity if the vacuum chambers were longer and could have a larger diameter, allowing for a stronger scalar field to build up. This could be realized with a split-crystal interferometer, cf. Fig.~\ref{fig:setup}~(b), which is currently under construction at our neutron interferometry station S18 at the Institut Laue-Langevin (ILL) in Grenoble. We envisage a beam separation of 10 cm and a chamber length of 0.5 m. A proof-of-principle experiment with a small split-crystal interferometer was already carried out ~\cite{lemmel2022neutron}. 

In either setup the incident neutron beam with mean wavelength $\lambda_{\rm n}=2.72$ \AA \, ($\delta\lambda_{\rm n}/\lambda_{\rm n}\sim0.043$, Bragg angle 45$^\circ$) and $4 \times 8 \,\mathrm{mm^{2}}$ beam cross section is split and passes the air or vacuum chamber respectively. After recombination at the last interferometer plate the intensity in the exiting beams (O in forward and H in refracted direction) is measured by detectors with an efficiency above 99 percent.
The intensity oscillates between the two exits as function of the phase. By rotating the phase shifter flag a sinusoidal fringe pattern is recorded. An additional phase created by the vacuum chambers can then be determined by a shift of the fringes.

 
One chamber, labelled ``vacuum'' in Fig.~\ref{fig:setup}, is evacuated to let the scalar field build up while the other chamber, labelled ``air'', is filled with some gas to suppress the field. This configuration creates primarily a phase shift due to the gas index of refraction, which is in the order of 100\textdegree~to 1000\textdegree~at ambient pressure. By using a gas with a well-known composition, e.g. pure Helium, this phase shift could be calculated and corrected for to high precision. However, in our experiment we reduced the pressure in the air chamber from ambient 1000 mbar to 0.01 mbar. While this pressure is high enough to suppress the scalar field, it is low enough to reduce the gas phase shift to below detection limit. The pressure in the vacuum chamber is still lower by a few orders of magnitude. In profile mode, the vacuum chambers are moved sidewards to vary the distance between the neutron beam and the side walls, thereby probing the shape of the scalar field within the chamber. 



\section{Derivation of dilaton constraints}\label{sec:IV}

\begin{figure}[h]
\centering
\includegraphics[width=0.48\textwidth]
{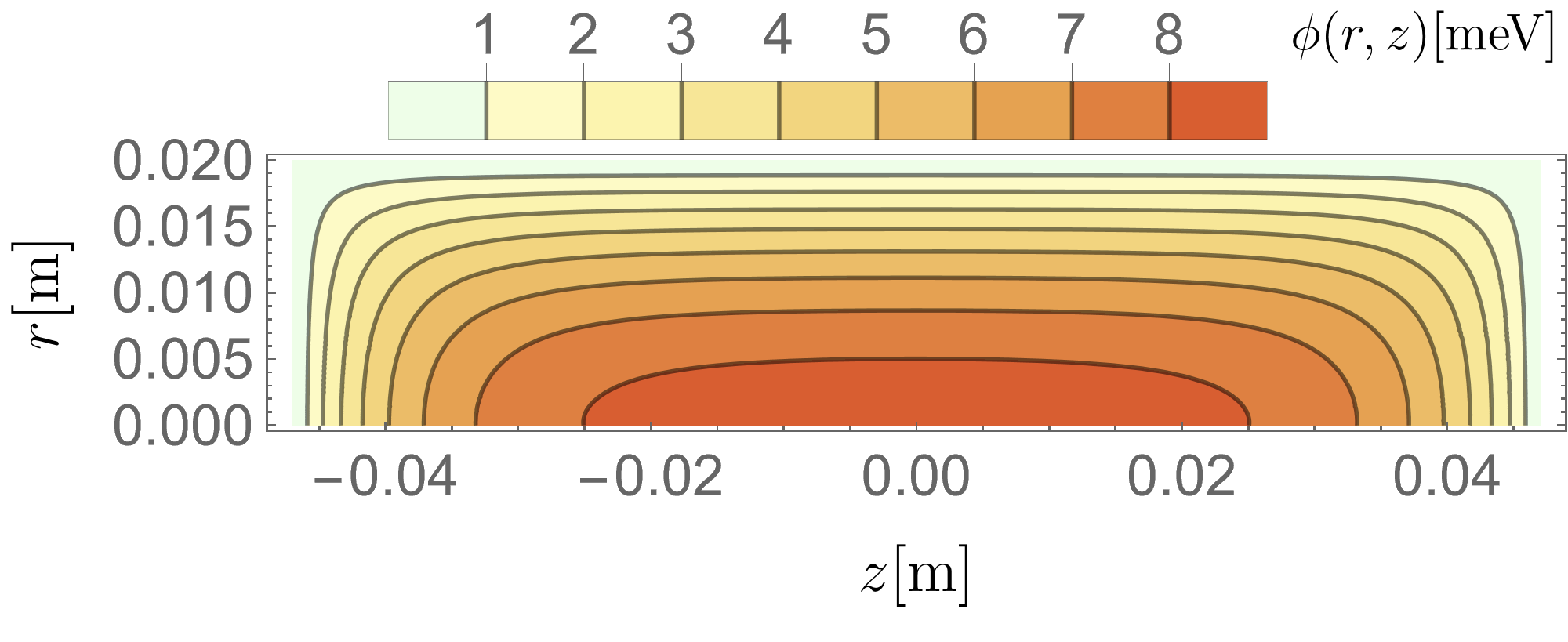}
\caption{Simulated dilaton field for $V_0 = 10$ MeV$^4$, $A_2 = 10^{37}$ and $\lambda_D = 10^{-9}$ with $\rho_V= 7.08 \times 10^{-17}$ MeV$^4$. \label{fig:DField}}
\end{figure}

\begin{figure*}[t]
\centering
\includegraphics[width=\textwidth]
{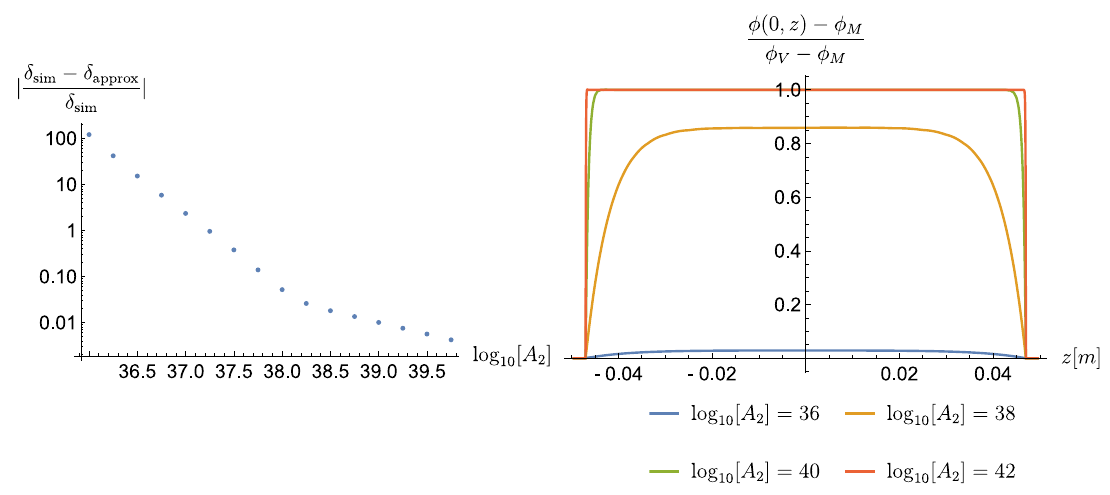}
\caption{The plots are computed for $\lambda_D=10^{-9}$ and $V_0=10$ MeV$^4$. \textit{Left:} The relative error of the approximation as a function of $A_2$ is plotted. \textit{Right:} Normalized field profiles along the $z$ axis for $r=0$. The value 1 corresponds to $\phi(0,z)=\phi_V$, while 0 corresponds to $\phi(0,z)=\phi_M$. The strong slopes occur at the chamber walls. \label{fig:Convergence}}
\end{figure*}

\begin{figure}[h]
\centering
\includegraphics[width=0.45\textwidth]
{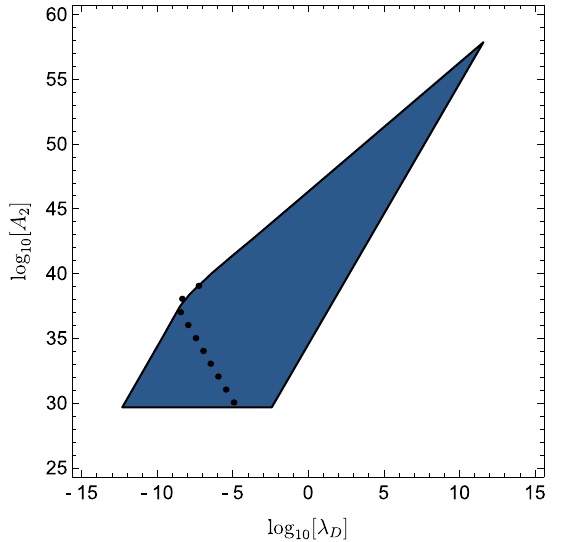}
\caption{The blue area corresponds to constraints given by the short-range approximation Eq.~\ref{approxS}. The bullet points are on the real contour of the constrained region obtained from simulations with $V_0=10$ MeV$^4$. \label{fig:Example}}
\end{figure}

\begin{figure*}[t]
\centering
\includegraphics[width=1\textwidth]
{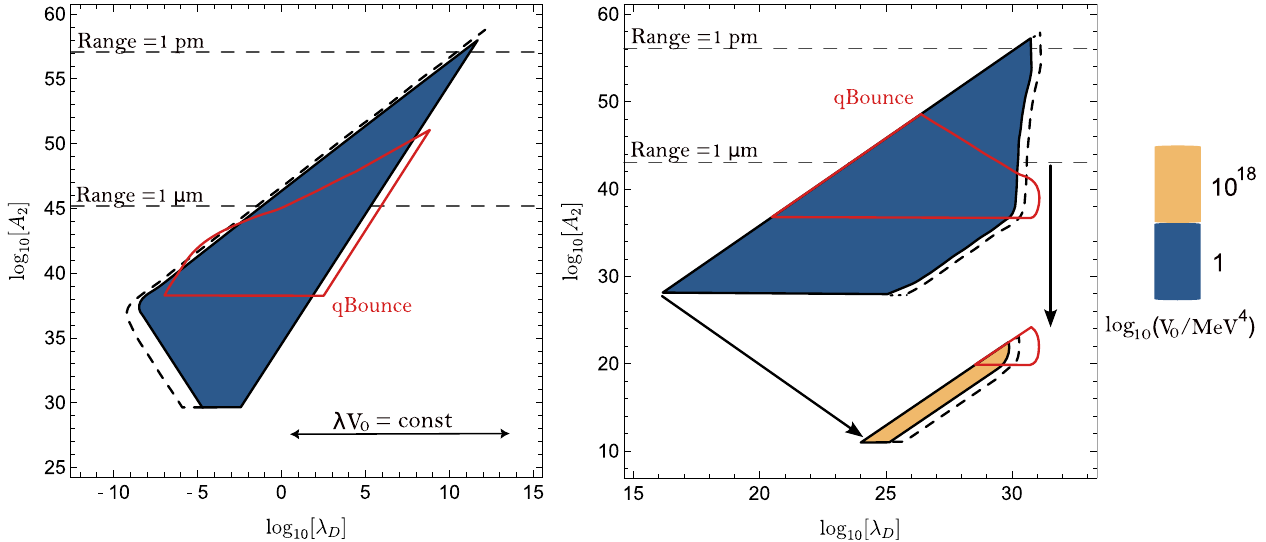}
\caption{Here, the constrained regions obtained from neutron interferometry for the dilaton field are plotted in color. The red lines correspond to existing constraints from \qbounce~\cite{Fischer:2023koa}, assuming a radius of 0.5 fm for the neutron as well. Dashed lines correspond to improvements expected for the geometry of a split interferometer (cylindrical vacuum chamber with length of 0.5 meters and radius 4.75 cm). \textit{Left:} In this part of the parameter space, the constrained area does not change its shape for increasing $V_0$ but only shifts towards lower $\lambda_D$. \textit{Right:} The constrained areas shift systematically towards lower values of $A_2$ for increasing $V_0$ without changing their shape but become smaller due to a cut-off. This is indicated by the arrows. The lower cut-off is a short-range cut-off to ensure that the field decays to its potential minimum inside the cylinder shell, the remaining cuts are due to the condition given in Eq.~\ref{cutoff}. \label{fig:FullNeut} The dilaton range is plotted for $V_0=10\ \text{MeV}^4$ and $\rho_V=7.08\times 10^{-17}$  MeV$^4$.}
\end{figure*}

The extremely high computational cost of solving the non-linear differential equation for many parameter values requires to lower the amount of simulations needed to find the constrained parameter volume (an example of a simulated dilaton field is given in Fig.~\ref{fig:DField}). While constraints from the profile mode are stronger for long interaction ranges where the field varies significantly throughout the chamber, the pressure mode is superior at probing short interaction ranges. Fields with very short interaction ranges are more difficult to simulate, since the slopes close to the chamber walls get arbitrarily steep requiring an ever finer discretization.
In this case, however,  the following approximation is highly accurate inside the vacuum chamber (obvious analogous expressions for the air chamber are omitted)
\begin{align}
    \int_{0}^Ldz\left(\phi^2(r = 0,z) - \phi^2_{\text{Air}}\right) \simeq L \left(\phi^2_V - \phi^2_{\text{Air}}\right),
\end{align}
since the field close to the chamber walls reaches its potential minimum $\phi_V$ quickly. For parameter values corresponding to this case, the profile mode cannot set constraints anymore, because the field effectively looses its position dependence. To illustrate this point we define
\setlength{\abovedisplayskip}{10pt} 
\setlength{\belowdisplayskip}{10pt}
\begin{align}
\delta_{\text{sim}} &:= - \mathfrak{Q}_D\frac{A_2m_n^2}{2k_0m^2_{\text{Pl}}} \int_{0}^Ldz \left(\phi^2(0,z)-\phi^2_{\text{Air}}\right), \\
\delta_{\text{approx}} &:= - \mathfrak{Q}_D\frac{A_2m_n^2}{2k_0m^2_{\text{Pl}}}\,L \left(\phi^2_V-\phi^2_{\text{Air}}\right). \label{approxS}
\end{align}
\setlength{\abovedisplayskip}{\medskipamount} 
\setlength{\belowdisplayskip}{\medskipamount} 
Here, $\delta_{\text{sim}}$ is the actual phase shift computed from a simulation for a neutron propagating through the center of the chamber ($r=0$), while $\delta_{\text{approx}}$ is an approximation that is valid only for very short-ranged fields.
For the dilaton field, small interaction ranges correspond to large values of $A_2$. We demonstrate in Fig.~\ref{fig:Convergence} that the error in the approximation becomes continuously smaller for larger values of $A_2$ and that the field takes on its VEV for an ever larger region inside the cylinder.
The constraints were computed as follows. For fixed $V_0$ and beginning from the smallest allowed value of $A_2$ (this comes from the long-range cut-off to ensure that the field decays to $\phi_M$ inside the shell of the vacuum chamber) a search was performed for the contour of the constrained region along the $\lambda_D$-axis with a step width of 0.1 in a logarithmic plot. Next, $A_2$ was increased  by a factor of 10 and the procedure repeated until the regime was entered, where the pressure mode dominates and the approximation becomes indistinguishable from the simulations. For even larger values of $A_2$, only the approximation was used in order to compute the rest of the constrained region. Next, the points just at the edge of the constrained region were connected. An example of this procedure is shown in Fig.~\ref{fig:Example}. 
For still larger values of $A_2$, a limit is reached where the necessary resolution of the finite element grid becomes so high that the field can practically no longer be computed. Therefore, the approximation solves multiple numerical challenges. The full constraints combining the pressure and profile mode are shown in Fig.~\ref{fig:FullNeut} together with already existing constraints from gravity resonance spectroscopy (\qbounce). 
Lunar Laser Ranging constraints cover much smaller values of $A_2$ than tabletop experiments and, therefore, these constraints are not shown~\cite{Fischer:2023koa}.

\section{Derivation of symmetron constraints}\label{sec:V}

The effective potential of the symmetron field \cite{khoury4525symmetron}
\begin{align}
     V_{\text{eff}}(\phi) = \frac{1}{2}\left(\frac{\rho}{M^2}-\mu^2\right)\phi^2 + \frac{\lambda_S}{4}\, \phi^4\>,
 \end{align}
 allows for spontaneous symmetry breaking.
In regions of high density, for which $\frac{\rho}{M^2}-\mu^2>0$ holds, there is only one real minimum of the effective potential at $\phi = 0$. The field is in its symmetric phase and fifth forces vanish.  However, in regions of low densities, where $\frac{\rho}{M^2}-\mu^2<0$ holds, the field is in its symmetry-broken phase and obtains a non-vanishing VEV inducing a fifth force. Furthermore, due to the $\phi \rightarrow - \phi$ symmetry of the potential the differential equation Eq.~\ref{eqn} can have more than one solution (see~\cite{Brax:2017hna,Pitschmann:2020ejb}). This property of the field is elucidated in the next subsection. For the calculation of constraints, we exclusively used solutions with a single node, which correspond to the lowest energy solution in general (for an example see Fig.~\ref{fig:P1}). We simulated the field numerically and found that for large enough $\mu$-values no solution exists. 
This is in agreement with~\cite{Upadhye:2012rc,Brax:2017hna}, where it was found that for large enough values of $\mu$ no solution can exist between two infinitely extended mirrors. Therefore, with the given dimensions of the vacuum and air chambers for symmetron ranges of  $\sim$1 mm and larger no field solution can exist anymore. 

\begin{figure}[h]
\centering
\includegraphics[width=0.5\textwidth]
{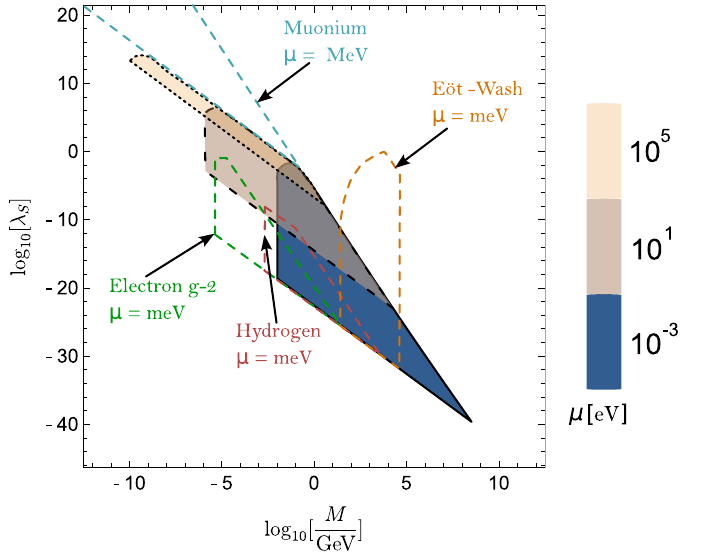}
\caption{The constrained volume for the parameters of the symmetron field obtained from neutron interferometry is shown in color, as well as existing constraints from atomic precision tests  and the Eöt-Wash experiment. The lower cut-off comes from the condition given in Eq.~\ref{cutoff}. At the left edge the symmetry of the potential is restored inside the vacuum region and the field vanishes everywhere.}
\label{fig:NeutSym}
\end{figure}

\begin{figure}[h]
\centering
\includegraphics[width=0.5\textwidth]
{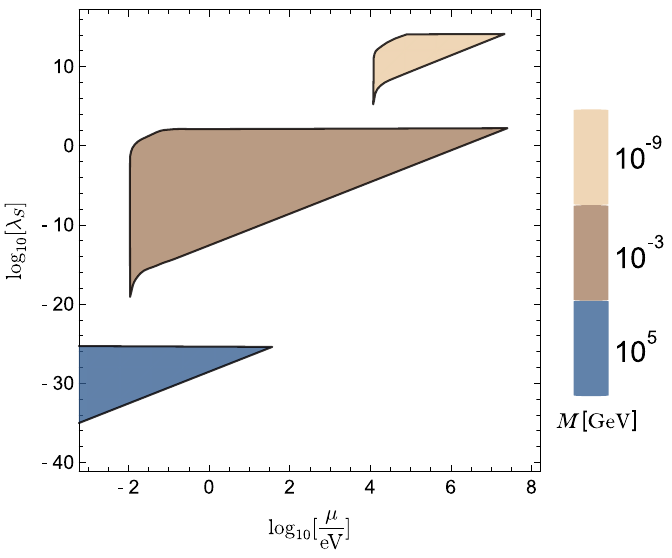}
\caption{Here, the constrained volume for the parameters of the symmetron field obtained from neutron interferometry for several fixed values of $M$ is depicted. \label{fig:NeutSym2}}
\end{figure}

This natural restriction of the symmetron field limits the parameter space to be probed to short-ranged fields. It was found that using the approximation $\phi \simeq \phi_V$ in the integral for the phase shift is accurate enough in the whole parameter space where new constraints can be set. In Fig.~\ref{fig:NeutSym} constraints obtained from neutron interferometry are plotted alongside already existing constraints published in~\cite{upadhye2013symmetron,brax2023screened}. 
The improvements of the split interferometer are too small to be visible in this log-log plot. For an alternative presentation with $\mu$ varying continuously see Fig.~\ref{fig:NeutSym2}.

\subsection{Domain walls and multiple node solutions}

\begin{figure}[h]
\centering
\includegraphics[width=0.48\textwidth]
{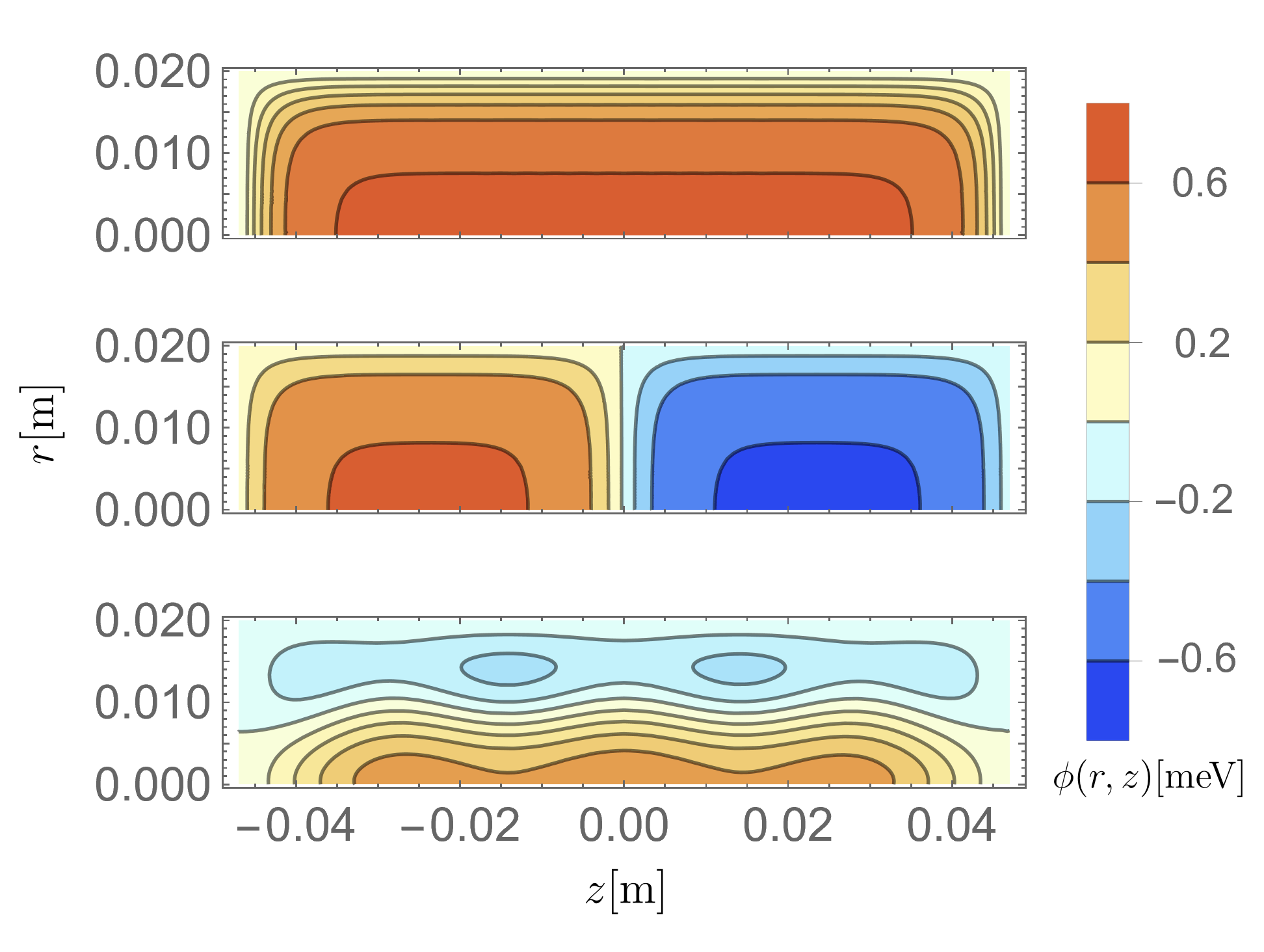}
\caption{The field profiles of the three solutions from Table~\ref{tab:onecolumn} are shown. For the given parameters $\phi_V = 0.61$ meV holds. \label{fig:P1}}
\end{figure}

Since in the symmetry-broken phase there are two distinct VEVs, the possibility arises that the symmetron settles to the positive VEV in some regions of space and to the negative VEV in others. Regions where the field takes on either VEV are called domains, while the boundaries connecting these domains are the domain walls (see e.g.~\cite{vachaspati2007kinks}).  While the latter might be unstable on cosmological time scales~\cite{christiansen2023asevolution}, it has been suggested that they might lead to observable consequences for ultra-cold neutrons. This is because neutrons would be accelerated towards the domain wall, resulting in a deflection angle~\cite{llinares2019detecting}. Furthermore, by draining gas in a vacuum chamber and allowing the field to relax to either VEV, such domain walls might deliberately be induced~\cite{clements2023detecting}.
A search for static domain walls inside a realistic cylindrical vacuum chamber was performed by trying a large number of plausible initial guesses for Newton's method in solving Eq.~\ref{numeric}. The spectrum of solutions found for a fixed parameter combination is shown in Fig.~\ref{fig:P1}. 
 The top solution has only one domain and one local extremum at the center of the chamber, while the solution in the middle has two domains with the field at its positive VEV in the chamber region $z<0$ and at its negative VEV for $z>0$. We note that the latter solution displays the anti-symmetry $\phi(r,-z) = - \phi(r,z)$, which shows that in the symmetry-broken phase the field does not have to obey the symmetries of the chamber. Instead of a solution having three domains along the $z$-axis, the solution displayed at the bottom has been found. While it does display multiple local extrema along either axis, none of them approaches any VEV too closely. This implies that static solutions can exist without the need for the field to take on its VEV anywhere inside the vacuum chamber. Despite using many more seeds for Newton's method, no further solutions have been obtained. However, we suspect that also solutions without cylinder symmetry exist, since the symmetron field solutions do not necessarily obey the symmetry of its environment. Finding such solutions would require solving the full 3D differential equation, which is beyond the scope of this article. We further suspect that solutions containing more domains can exist only for larger parameter values of $\mu$, where the field can have more curvature. This behavior, including the absence of any solutions for too small values of $\mu$,  has already been observed in~\cite{Brax:2017hna,Pitschmann:2020ejb} for the case of a 1-dimensional setup containing two parallel mirrors. The energies of these field solutions and their interaction with the matter density can be obtained by the Hamilton density 
  \begin{align}
      \mathcal{H} &=  \frac{1}{2}\left(\vec{\nabla}\phi\right)^2 + V_{\text{eff}}(\phi)\nonumber\\
      &= \frac{1}{2} \left\{\left(\frac{\partial \phi}{\partial r}\right)^2 + \left(\frac{\partial \phi}{\partial z}\right)^2\right\} + V_{\text{eff}}(\phi)\>,
  \end{align}
 where cylinder symmetry is assumed such that derivatives with respect to $\varphi$ vanish. This results in the corresponding energy $E$ inside the cylinder
 \begin{align}
     E = \int\limits_{\text{cylinder}} \hspace{-3mm}d^3 x\,\mathcal{H}(\Vec{x}) = 2 \pi \int_{0}^R dr\,r\int_{-L/2}^{L/2} dz\, \mathcal{H}(r,z)\>,
 \end{align}
where $R$ and $L$ refer to the radius and length of the cylinder, respectively. The obtained energies are listed in Table~\ref{tab:onecolumn}.
\begin{table}[h]  
\centering
\addtolength{\tabcolsep}{2pt}
\renewcommand{\arraystretch}{1.4}
\begin{tabular}{|c||c|}
  \hline
  Mode & $E$ [eV]   \\
  \hline\hline
   0 & -2.097 \\
   \hline
   1 & -1.710  \\
   \hline
   2 & -0.06\\
  \hline
\end{tabular}
\caption{Energies $E$ are given corresponding to the three obtained solutions with the mode number enumerating the solutions with increasing energy. Hereby, the following numerical values have been used $\rho_M= 1.16 \times 10^{-5}$ MeV$^4$, $\rho_V= 7.08 \times 10^{-17}$ MeV$^4$, $\lambda_S=10^{-2}$, $M=10^{3}$ GeV, and $\mu = 6.1\times10^{-5}$ eV.}
    \label{tab:onecolumn}
\end{table}

\section{Comments on chameleon constraints}\label{sec:VI}

For the chameleon field, each value of $n$ in the potential corresponds to a separate model. Restricting the analysis to parameters that have been studied frequently, no constraints improving on already existing ones are obtained for $n=-4$  (in this case $V_C(\phi)$ is replaced by $V_C(\phi)= \lambda_C \phi^4$, with a dimensionlass coupling constant $\lambda_C$) and $n = 1$ with varying $\Lambda$. Also setting $\Lambda = 2.4$ meV to the dark energy scale 
and varying $n$ for small values does not provide more stringent constraints than those available. Therefore, no exclusion plots are provided herein. For comparison to existing constraints use has been made of Ref.~\cite{burrage2018tests}.

\section{Discussion and conclusion}\label{sec:VII}

Since very short ranges of scalar fields induce short-range fifth forces, classical experiments eventually are no longer able to detect or probe scalar fields in this regime. However, for neutron interferometry very short ranges of scalar fields lead effectively to a constant potential shift inside a vacuum chamber after a steep gradient resulting in a phase shift. The relative phase shift due to the different potential shifts induced in the two chambers can still be measured even when classical experiments would fail to detect any force. For this reason, constraints were obtained for extremely large values of $A_2$ and $\mu$ (which corresponds to small ranges) for the dilaton and symmetron field. However, since the neutron has a finite extent, eventually the screening of the neutron itself sets in and prohibits probing arbitrarily small ranges. 

While a split-crystal interferometer opens up the possibility to increase the chamber length by a small factor, this does not yield a large improvement. However, lowering the vacuum pressure and hence the vacuum density by a few orders of magnitude (the current setup uses a vacuum pressure of $10^{-4}$ mbar or higher, which might eventually be lowered to $10^{-9}$ mbar) seems to provide a more powerful way to probe screened scalar fields, since they generically are less suppressed in regions of lower mass density. Finally, it was found that for the symmetron inside cylindrical vacuum chambers in general several distinct static field solutions exist.

\section{Acknowledgments}

We thank Hartmut Abele and Tobias Jenke for fruitful discussions. 
This article was supported by the Austrian Science Fund (FWF): P 34240-N, and is based upon work from COST Action COSMIC WISPers CA21106,
supported by COST (European Cooperation in Science and Technology).

\appendix

\section{Numerical solutions of the differential equations}\label{sec:AppI}

In order to solve 
\begin{align}
    \Delta \phi (\vec{x}) = V_{\text{eff}, \phi} (\phi(\vec{x}); \rho(\vec{x}))\>,
\end{align}
Newton's method has been applied on the function space. This method has already previously been used to simulate chameleon fields~\cite{briddon2021selcie}. Starting from an initial guess function $\phi^{(0)}(\vec{x})$ for the field, the differential equation is expanded to first order around that guess function. The solution serves as an improved guess. This results in a sequence of functions $\phi^{(0)}$, ...$\phi^{(n)}$, ... with $\phi^{(n)}$ defined as the solution of the $linear$ differential equation
\begin{align}
    &\Delta \phi^{(n)} (\vec{x}) = V_{\text{eff}, \phi} (\phi^{(n-1)}(\vec{x});\rho(\vec{x})) \nonumber\\ 
    &+V_{\text{eff}, \phi \phi} (\phi^{(n-1)}(\vec{x});\rho(\vec{x}))\left(\phi^{(n)}-\phi^{(n-1)}\right).
\end{align}
These linear differential equations were solved repeatedly with the finite element method (see e.g.~\cite{langtangen2019introduction}) until 
\begin{align}
    \frac{||\phi^{(n)}-\phi^{(n-1)}||_2}{||\phi^{(n-1)}||_2}<\epsilon\>,
\end{align}
for some small value of $\epsilon$ (which was at least set to $10^{-10}$, but much smaller for extreme parameters). To find the one node solution, the effective potential was first expanded around its minimum
\begin{align}
    &\left[\frac{1}{r}\frac{\partial}{\partial r}\left(r\frac{\partial}{\partial r}\right)  + \frac{\partial^2}{\partial z^2}\right]\phi(r,z)=\nonumber\\
     &\qquad V_{\text{eff}, \phi \phi}(\phi_{\rho(r,z)})\left(\phi(r,z)-\phi_{\rho(r,z)}\right),
\end{align}
which resulted in a linear differential equation that can be solved explicitly. Then, the solution was used as an initial guess for Newton's method to solve the full non-linear differential equation. The multiple node solutions for the symmetron field were found by trying a large number of physically plausible seeds.

\section{Derivation of the phase shift formula}\label{sec:AppII}

The derivation below largely follows~\cite{greenberger1979coherence} (see also~\cite{MarioHabil}).  Starting from the free Schrödinger equation for the neutron
\begin{align}
\hat H_0 \phi_0 &= -\frac{1}{2m_n}\,\Delta \phi_0 \nonumber\\
&= E_0 \phi_0\>,
\end{align}
where $\phi_0\propto e^{i \vec{k}\cdot\vec{x}}$, the ansatz 
\begin{align}
\phi &= \phi_0 \chi\>, \nonumber\\
 E&=E_0\>,
\end{align}
is performed. The full Schrödinger equation reads
\begin{align}
\big(\hat H_0+U\big)\phi = E\phi\>.
\end{align}
A straight forward calculation gives
\begin{align}\label{eq:App1}
\phi_0 \Delta \chi + 2i \phi_0 \vec{k}_0\cdot \vec{\nabla}\chi = 2m_nU\phi_0 \chi\>.
\end{align}
The semi-classical limit is given by $|\vec{\nabla} \chi| \ll k_0 |\chi|$, which allows to neglect the first term. Defining $k_0 := |\vec{k}_0| = 2\pi/\lambda$ and  the length parameter $s$ along the direction of $\vec{k}_0$ gives $\vec{k}_0\cdot \vec{\nabla} = k_0\,\frac{d}{ds}$ and 
\begin{align}
\chi &= e^{- i\frac{m_n}{k_0} \int ds\,U}\>. \label{wavefunc}
\end{align}
The resultant phase shift is therefore given by
\begin{align}
\delta\varphi = - \frac{m_n}{k_0} \int ds\,U\>,
\end{align}
and for neutrons propagating through a cylindrical cavity with length $L$  along the symmetry $z$-axis at constant radius $r_c$ 
\begin{align}
\delta\varphi = - \frac{m_n}{k_0} \int_0^L dz\, U(r_c,z)\>.
\end{align}
The relative phase shift between vacuum chamber and air chamber takes the form
\begin{align}
\delta\varphi_{\text{relative}} = - \frac{m_n}{k_0} \int_0^L dz\left(U_{\text{Vac}}(r_{c_1},z)- U_{\text{Air}}(r_{c_2},z)\right).
\end{align}

Next, the condition for the validity of the semiclassical limit will be derived. From 
\begin{align}\label{eq:App2}
\vec{\nabla}\chi =- i\frac{m_n}{k_0}\, U(r_c,z)\,\chi\,\vec{e}_z\>,
\end{align}
follows with $|\chi|=1$
\begin{align}
|\vec{\nabla}\chi| =\frac{m_n}{k_0}\, |U(r_c,z)|\>.
\end{align}
Hence, the condition for the validity of the semiclassical limit becomes
\begin{align}
\frac{|\vec{\nabla}\chi|}{k_0} = \frac{m_n}{k_0^2} \, |U(r_c,z)| \ll 1\>.\label{semic}
\end{align}

Finally, the derivation assumes that outside both chambers, i.e. in air, the perturbation potential $U=0$. However, scalar fields in air typically take on a non-vanishing value. For all parameter values where constraints can be set, however, the ranges of the fields in air are extremely small compared to the dimensions of the experimental setup. Therefore, the field is to a very good approximation constant outside the chambers, having its air expectation value. In order to comply with the assumption $U=0$ outside the chambers, we have to define the potentials for the dilaton (D), symmetron (S) and chameleon (C) perturbation potentials in the following way
\begin{align}
U_D(\vec{x}) &:= \frac{A_2 m_n}{2 m^2_{\text{Pl}}}\left(\phi^2(\vec{x})-\phi^2_{\text{Air}}\right), \\
U_S(\vec{x}) &:= \frac{m_n}{2 M^2}\left(\phi^2(\vec{x})-\phi^2_{\text{Air}}\right), \\
U_C(\vec{x}) &:=  \frac{m_n}{M_c}\left(\phi(\vec{x})-\phi_{\text{Air}}\right).
\end{align}

The potentials above are only valid in the limit where the neutron can be considered as a test particle in a scalar field background rather than as a source of the field, i.e.\,\,if the screening of the neutron can be neglected. In order to take its screening into account, we multiply the above potentials with a screening charge
\begin{align}
    U_X(\vec{x}) \rightarrow \mathfrak{Q}_X U_X(\vec{x})\>,
\end{align}
with $X\in \{D, S, C\}$. For additional information on the screening charge of neutrons see Appendix A of Ref.~\cite{Brax:2017hna}. Therefore, the resulting conditions for the validity of the semiclassical limit are given by
\begin{align}
\mathfrak{Q}_D\frac{A_2 m_n^2}{2 m^2_{\text{Pl}}k_0^2}\left(\phi^2(\vec{x})-\phi^2_{\text{Air}}\right)&\ll 1\ (\text{dilaton})\>,\\
\mathfrak{Q}_S \frac{m_n^2}{2 M^2k_0^2}\left(\phi^2(\vec{x})-\phi^2_{\text{Air}}\right)&\ll 1\ (\text{symmetron})\>,\\
\mathfrak{Q}_C \frac{m_n^2}{M_ck_0^2}\left(\phi(\vec{x})-\phi_{\text{Air}}\right)&\ll 1\ (\text{chameleon})\>.
\end{align}
These conditions are  always fulfilled on those edges of the constrained areas, which do not come from imposed cut-offs.

\section{Screening charge for the chameleon field}\label{sec:AppIII}

The following derivation and physical reasoning closely follows \cite{Brax:2022uyh} and hence only key steps in the derivation of the screening charge for the chamleon field are provided. In order to compute the chameleon field of a spherical source, one has to solve
\begin{align}
    \frac{d^2\phi}{dr^2}+\frac{2}{r}\frac{d\phi}{dr} = V_{\text{eff},\phi}(\phi; \rho)\>.
\end{align}
Since the field outside the sphere approaches its asymptotic value $\phi_V$ and inside the sphere $\phi_S$, the effective potential is expanded around these values
\begin{align}
    \frac{d^2\phi}{dr^2}+\frac{2}{r}\frac{d\phi}{dr} &=\mu^2_S\left(\phi - \phi_S\right), &r < R\>, \nonumber\\
    \frac{d^2\phi}{dr^2}+\frac{2}{r}\frac{d\phi}{dr} &=\mu^2_V\left(\phi - \phi_V\right), &r \geq R\>,
\end{align}
where $R$ refers to the radius of the sphere, $\mu_S$ to the chameleon mass for the field taking its minimum value inside the sphere and $\mu_V$ to the chameleon mass for the field taking its VEV. The final solution outside the sphere is given by
\begin{align}
    \phi(r) = \phi_V - \mathfrak{Q}_C \frac{\mu^2_S R^3}{3} \left(\phi_V-\phi_S\right) \frac{e^{- \mu_V(r-R)}}{r}\>,
\end{align}
where $\mathfrak{Q}_C$ is the screening charge of the chameleon field defined by
\begin{align}
    \mathfrak{Q}_C := \frac{3}{\mu^2_S R^2} \frac{1-\frac{1}{\mu_S R}\, \text{tanh}(\mu_S R)}{1+\frac{\mu_V}{\mu_S}\,\text{tanh}(\mu_S R)}\>,
\end{align}
which describes the screening of the sphere. The limits
\begin{align}
  \mathfrak{Q}_C\rightarrow\begin{dcases}
    0\>, & \text{for ``screened" bodies with }\mu_S R \rightarrow \infty\>, \nonumber\\
    1\>, & \text{for ``unscreened" bodies with }\mu_S R \rightarrow 0\>,
  \end{dcases}
\end{align}
are obeyed.

\bibliography{refs}
 
\end{document}